\documentclass[twocolumn,aps,prl,superscriptaddress]{revtex4-1}
\usepackage{amssymb} 
\usepackage{amsmath,bm} 
\usepackage{graphicx} 
\usepackage[normalem]{ulem} 
\usepackage{multirow} 
\usepackage[colorlinks,linkcolor=blue,urlcolor=blue,citecolor=blue]{hyperref} 
\usepackage{lipsum} 
\usepackage[usenames, dvipsnames]{xcolor} 
\usepackage{tensor} 
\usepackage{isotope} 
\usepackage{amsmath} 
\allowdisplaybreaks[4] 

\renewcommand{\sout}{\bgroup \color{red} \ULdepth=-.5ex \ULset} 
\newcommand{\com}[1]{{\sf\color[rgb]{0,0,1}{#1}}}

\newcommand{\hyt}{$^3_\Lambda \text{H}$}

\newcommand{\hth}{$^3_\Lambda \text{H}/^3\text{He}$}
\begin{document} 

%\title{Quantum Mechanical Softening of the Hypertriton Transverse Momentum Spectrum in Heavy-Ion Collisions  }
\title{Softening of the Hypertriton Transverse Momentum Spectrum in Heavy-Ion Collisions  }

\author{Dai-Neng Liu}
%\email{dnliu17@fudan.edu.cn}
\affiliation{Key Laboratory of Nuclear Physics and Ion-beam Application~(MOE), Institute of Modern Physics, Fudan University, Shanghai $200433$, China} 
\affiliation{Shanghai Research Center for Theoretical Nuclear Physics, NSFC and Fudan University, Shanghai 200438, China}  

\author{Che Ming Ko} 
\email{ko@comp.tamu.edu} 
\affiliation{Cyclotron Institute and Department of Physics and Astronomy, Texas A\&M University, College Station, Texas 77843, USA} 

\author{Yu-Gang Ma} 
\email{mayugang@fudan.edu.cn} 
\affiliation{Key Laboratory of Nuclear Physics and Ion-beam Application~(MOE), Institute of Modern Physics, Fudan University, Shanghai $200433$, China} 
\affiliation{Shanghai Research Center for Theoretical Nuclear Physics, NSFC and Fudan University, Shanghai 200438, China} 
 
\author{Francesco Mazzaschi} 
\affiliation{Dipartimento di Fisica dell’Università and Sezione INFN, Turin, Italy} 

 \author{Maximiliano Puccio} 
\affiliation{Experimental Physics Department, CERN, CH-1211 Geneve 23, Switzerland} 
 
\author{Qi-Ye Shou}
\affiliation{Key Laboratory of Nuclear Physics and Ion-beam Application~(MOE), Institute of Modern Physics, Fudan University, Shanghai $200433$, China} 
\affiliation{Shanghai Research Center for Theoretical Nuclear Physics, NSFC and Fudan University, Shanghai 200438, China}  

\author{Kai-Jia Sun} 
\email{kjsun@fudan.edu.cn} 
\affiliation{Key Laboratory of Nuclear Physics and Ion-beam Application~(MOE), Institute of Modern Physics, Fudan University, Shanghai $200433$, China} 
\affiliation{Shanghai Research Center for Theoretical Nuclear Physics, NSFC and Fudan University, Shanghai 200438, China} 

\author{Yuan-Zhe Wang} 
\affiliation{Key Laboratory of Nuclear Physics and Ion-beam Application~(MOE), Institute of Modern Physics, Fudan University, Shanghai $200433$, China} 
\affiliation{Shanghai Research Center for Theoretical Nuclear Physics, NSFC and Fudan University, Shanghai 200438, China} 

\date{\today}

%key: anti-matter, Y-N interaction, and weakly bound state production with different structures.
\begin{abstract}
Understanding the properties of hypernuclei helps to constrain the interaction between hyperon and nucleon, which is known to play an essential role in determining the properties of neutron stars.  Experimental measurements have suggested that the hypertriton (\hyt), the lightest hypernucleus, exhibits a halo structure with a deuteron core encircled by a $\Lambda$ hyperon at a distance of about 10 fm.  This large $\Lambda-d$ distance in \hyt ~wave function is found to cause a suppressed \hyt ~yield and a softening of its transverse momentum ($p_T$) spectrum in relativistic heavy-ion collisions. Within the coalescence model based on nucleons and $\Lambda$ hyperons from a microscopic hybrid hydro model with a hadronic afterburner for nuclear cluster production in Pb-Pb collisions at $\sqrt{s_{NN}}$= 5.02 TeV, we show how this softening of the hypertriton $p_T$ spectrum appears and leads to a  smaller  mean $p_T$ for \hyt ~than for helium-3 ($^3$He). The latter  is opposite to the  predictions from the blast-wave model which assumes that \hyt~and $^3$He are thermally produced at the kinetic freeze-out of heavy-ion collisions. The discovered quantum mechanical softening of the (anti-)hypertriton spectrum can be experimentally tested in relativistic heavy-ion collisions at different collision energies and centralities and used to obtain valuable insights to the mechanisms for light (hyper-)nuclei production in these collisions.
\end{abstract}

%\pacs{12.38.Mh, 5.75.Ld, 25.75.-q, 24.10.Lx}
\maketitle

%1.1 Y-N interaction and hypernuclei
\emph{Introduction.}{\bf ---}
Observations of anti-nuclei such as anti-deuteron ($\bar{d}$), anti-helium ($^3\overline{\text{He}}$ and $^4\overline{\text{He}}$), and anti-hypertriton ($^3_{\bar{\Lambda}}\overline{\text{H}}$) have been  reported in many heavy-ion collision experiments~\cite{STARSc328,STARNt473,ALICE:2015rey,STARNtP16,Chen:2018tnh,Braun-Munzinger:2018hat, ALICE:2022veq}. Recent experiments at the Relativistic Heavy Ion Collider (RHIC)~\cite{STAR:2023fbc} and the Large Hadron Collider (LHC)~\cite{ALICE:QM2023hyper} have even suggested possible detection of heavier anti-hypernuclei, such as $^4_{\bar{\Lambda}}\overline{\text{H}}$ and $^4_{\bar{\Lambda}}\overline{\text{He}}$.  The detection of these bound nuclei and the  study of their properties \cite{Gal:2016boi,STAR_pp, ALICE:2022sco,ALICE:2019vlx,flow} are important in the search for the signals of phase transitions in the strongly interacting matter created in relativistic heavy-ion collisions~\cite{Sun:2017xrx,Sun:2018jhg,Sun:2020zxy,Shuryak:2018lgd,Shuryak:2019ikv,STAR:2022hbp,KoCM}  and in the dark matter detection in space~\cite{Blum:2017qnn,vonDoetinchem:2020vbj, ALICE:2022zuz} as well as for understanding the fundamental CPT theorem in quantum field theory~\cite{ALICE:2015rey,STARNtP16}.  Also, understanding the properties of hypernuclei  helps to constrain the hyperon-nucleon ($Y\text{-}N$) interaction~\cite{Chen:2023mel, Ma_NST, Ma_EPJC}, which is essential for determining the  structure of compact astrophysical objects like the neutron stars~\cite{Gal:2016boi}. 

Various theoretical models based on different assumptions have been used to describe (anti-)nuclei production in nuclear reactions.  These include the statistical hadronization model (SHM)~\cite{AndNt561},
the coalescence model~\cite{SchPRC59,Sun:2018mqq,Zhang:2018euf,Bellini:2020cbj}, and the kinetic or transport approach~\cite{DanNPA533,OhPRC76,OhPRC80,OliPRC99,Sun:2022xjr}.  While both (anti-)nuclei disintegration and regeneration during the evolution of the hadronic matter are included in the kinetic approach, only (anti-)nuclei production is considered in the statistical model and the coalescence model by assuming, respectively, that they are thermally produced from the hadronization of produced quark-gluon plasma (QGP) and formed from the recombination of nucleons at the kinetic freeze-out. Another difference between the statistical and the coalescence model is the role of light cluster wave function in its production.  While the wave function is essential for forming light clusters in the coalescence model, it is not relevant in the SHM. With the hypertriton characterized by a halo structure with a deuteron core surrounded by a $\Lambda$ hyperon, it provides an excellent tool for differentiating between these two models. If the coalescence model proves to be the production mechanism of light nuclei and hypernuclei in nuclear collisions, this will open a new way for investigating their structures and internal wave functions.

According to recent STAR and ALICE measurements, the \hyt~has a lifetime close to that of a free $\Lambda$~\cite{STAR:2021orx, ALICE:2022sco, ALICE:2019vlx} and a $\Lambda$ separation energy ($B_\Lambda$) of merely a few hundreds keV.  While the  $B_\Lambda$ reported by the ALICE Collaboration is $0.102\pm 0.063\text{(stat.)}\pm 0.067\text{(syst.)} $ MeV ~\cite{ALICE:2022sco},  it is  $0.41\pm 0.12\text{(stat.)}\pm 0.11\text{(syst.)}$ MeV according to the measurement of the STAR Collaboration~\cite{STARNtP16, Liu:2019mlm}. The world average $B_\Lambda$ is, however, $0.164 \pm 0.043$ MeV~\cite{ Juric:1973zq, KEYES1970, Chaudhari1968, Mayeur1966ADO, ammar1962, cobis_simplest_1997, STARNtP16, ALICE:2022sco, HypernuclearDataBase}, corresponding to a very large $\Lambda-d$ distance of $\sqrt{\langle r^2_{\Lambda d}\rangle}\approx 9.5$ fm, which  suggests that the \hyt~has a halo structure with a deuteron core surrounded by a $\Lambda$ halo.  The halo structure is an intriguing quantum tunneling phenomenon observed in nuclei far from the $\beta$-stability line~\cite{Tanihata}.  Within the coalescence  model~\cite{Sun:2018mqq}, it has been demonstrated that such a large $\Lambda-d$ distance lead to a pronounced suppression of \hyt~production in collisions of small systems, a phenomenon that has been consistently observed in various ALICE measurements at the LHC~\cite{ALICE:2017xrp, ALICE:2019fee, ALICE:2020foi, ALICE:2021ovi, ALICE:2021mfm, ALICE:2021sdc}.  While the grand-canonical ensemble version of the statistical hadronization model fails to account for the suppression of light clusters in small systems, the canonical statistical model, which takes into account exact charge conservation, does predict a similar suppression seen in the experimental data~\cite{Vovchenko:2018fiy}.  

In the present study, we demonstrate that the large $\Lambda-d$ distance in \hyt ~wave function further results in a softening of its transverse momentum ($p_T$) spectrum in Pb+Pb collisions at $\sqrt{s_{NN}}=5.02$ TeV. This softening effect on the hypertriton spectrum occurs not only in non-central collisions but also in central collisions, where the canonical effects are expected to be insignificant. The quantum mechanical softening of the hypertriton spectrum in the coalescence model thus makes it a promising tool to distinguish it from the statistical hadronization model. 

%%%%%%%%%%%%%%%%%%%%%%%%%%%%%%%%%%%%%%%%%%%%%%%%%%%%
%\section{results and discussions}

\emph{Suppression of hypertriton production in peripheral collisions}{\bf ---} We adopt the hybrid model~\cite{Zhao:2021dka} of MUSIC+UrQMD+COAL to study  the production of (anti-)hypertriton  in Pb+Pb collisions at $\sqrt{s_{NN}}=5.02$~TeV.   After the evolution of the QGP produced in these collisions via the (3+1)-dimensional viscous hydrodynamic model MUSIC~\cite{PaqPRC93, SCPRC97, SCNST31} with the collision-geometric-based 3D initial conditions~\cite{Shen:2020jwv} and a crossover type of equation of state at finite density NEOS-BQS~\cite{Monnai:2019hkn}, hadrons are produced from a constant energy density hypersurface according to the Cooper-Frye formula~\cite{CooPRD10}. The subsequent hadronic rescatterings and decays are modeled by the UrQMD model~\cite{Bass:1998ca}. The (anti-)hypertriton yield in a collision event is then obtained from the coalescence (COAL) of (anti-)protons, (anti-)neutrons, and (anti-)$\Lambda$ hyperons at their kinetic freeze-out.   

In the coalescence model~\cite{SchPRC59, Sun:2018mqq} for light cluster production, the formation probability of a hypertriton from  a proton, a neutron, and a $\Lambda$ hyperon is given by the product of the statistical factor $g_{ht}=1/4$ for spin 1/2 proton, neutron, and $\Lambda$ hyperon to form a spin 1/2 hypertriton and the Wigner function of the hypertriton internal wave function, which we take as
\begin{eqnarray}
W_\text{ht}&=&8^2  e^{-\frac{\rho ^{2}}{\sigma_1 ^{2}} -\sigma_1 ^{2} k_{\rho }^{2}} e^{-\frac{\lambda ^{2}}{\sigma_2 ^{2}} -\sigma_2 ^{2} k_{\lambda }^{2}}.   \label{Eq:wig}
\end{eqnarray}
The relative coordinates and momenta in the above equation are defined by
\begin{eqnarray}
 {\bf \rho} &=&\frac{{\bf r}_{1} -{\bf r}_{2}}{\sqrt{2}},~ {\bf \lambda} =\sqrt{\frac{2}{3}}\left(\frac{m_{1} {\bf r}_{1} +m_{2} {\bf r}_{2}}{m_{1} +m_{2}} -{\bf r}_{3}\right), \\
 \label{Eq:coordinate}
k_\rho &=&\sqrt{2}\frac{({m_2\bf k}_{1} -m_1{\bf k}_{2})}{m_1+m_2},  \\
k_\lambda &=&  \sqrt{\frac{3}{2}} \frac{m_{3} ({\bf k}_{1} +{\bf k}_{2})-(m_1+m_2){\bf k}_3}{m_{1} +m_{2}+m_{3}}.
\end{eqnarray} 
Here, ${\bf r}_1$, ${\bf r}_2$, and ${\bf r}_3$ are the spatial coordinates of neutron, proton, and $\Lambda$ hyperon, respectively\com{,} at an equal time in their rest frame, and they are determined by propagating the constituent particles of momenta ${\bf k}_1$, ${\bf k}_2$, and ${\bf k}_3$ at earlier freeze-out times to the time of the last freeze-out one. The size parameter $\sigma_{1}$ in the Wigner function is related to the deuteron root-mean-squared radius $r_{{d}}$ by $\sigma_{1}  = \sqrt{4/3}~r_{d}\approx 2.26$ fm~\cite{Sun:2017ooe, Ropke:2008qk}. The size parameter $\sigma_2$  is  related to the $\Lambda-d$ distance by $\sigma_2=\frac{2}{3}\sqrt{\langle r_{\Lambda d}^2\rangle}$, and it can be  parameterized  as $\sigma_2\approx (2.15 (B_\Lambda/\text{MeV})^{-1/2}+1.23)~$fm, with details provided in the Appendix\ref{appendix}.   For the binding energies measured by the STAR and ALICE collaboration, we obtain $\sigma_2=5.45$ fm and $\sigma_2=7.96$ fm, respectively, while using the world averaged $B_\Lambda$ value gives $\sigma_2=6.52$ fm. For helium-3 production in the coalescence model, we use $\sigma_1=\sigma_2\approx 1.76$ fm \cite{Ropke:2008qk} as it has a spherical shape. 
\begin{figure}[!t]
    \centering
    \includegraphics[scale=0.4]{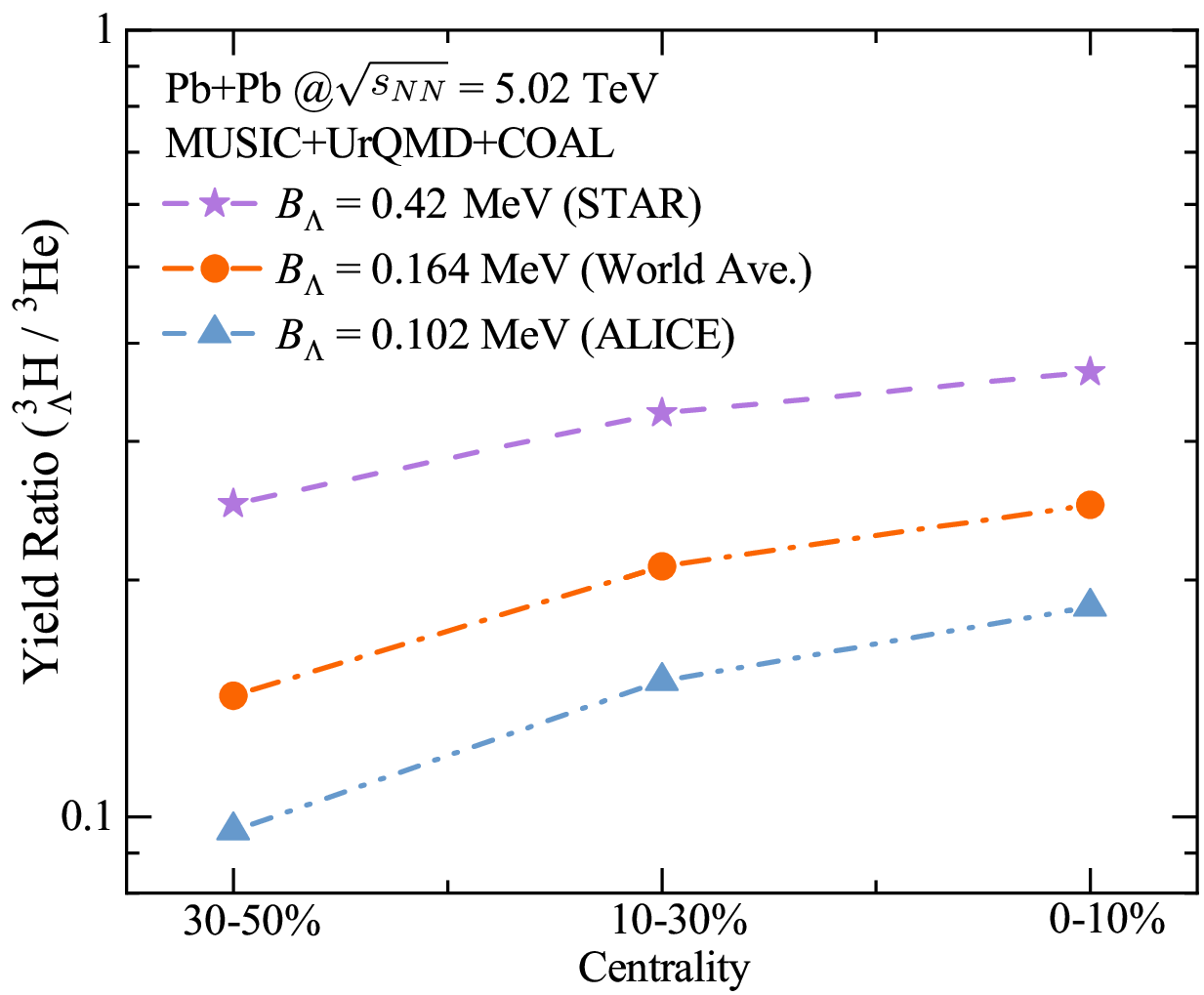}
    \caption{Collision centrality dependence of $_\Lambda^3$H/$^3$He yield ratio in Pb+Pb collisions at $\sqrt{s_{NN}}=5.02$ TeV for different $B_\Lambda$ values of 0.42 MeV (solid stars), 0.164 MeV (solid circles), and 0.102 MeV (solid triangles). }
    \label{fig:yield}
\end{figure} 

\begin{figure*}[!t]
    \centering
    \includegraphics[scale=1.45]{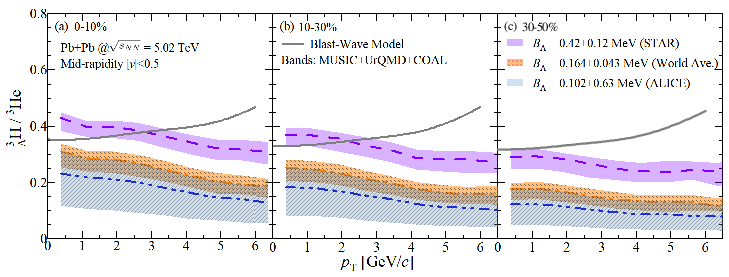}
    \caption{The yield ratio of hypertriton to helium-3 as a function of transverse momentum $p_T$ in Pb+Pb collisions at $\sqrt{s_{NN}} = 5.02$ TeV  for 0-10\% (a), 10-30\% (b), and 30-50\% (c) centralities. Results from the coalescence approach (MUSIC+UrQMD+COAL) and blast-wave model are denoted by shaded regions and solid lines, respectively. }
    \label{fig:pt_spectra}
\end{figure*}

We note that the statistical factor $g_{\rm ht}=1/4$ used in the coalescence model for proton, neutron, and $\Lambda$ hyperon to form a hypertriton does not include that due to their isospins.  As shown in Ref.~\cite{Sun:2018jhg}, the light nuclei numbers from such a coalescence model based on nucleons from a thermal source reduce to those from the thermal model when their binding energies are much smaller than the temperature and their radii are much larger than the size of the particle emission source.  The exclusion of the isospins in the coalescence model based on kinetically freeze-out nucleons in relativistic heavy-ion collisions is also supported by the results in Ref.~\cite{Sun:2022xjr} from the kinetic approach, which treats light nuclei as dynamic degrees of freedom by including their production and annihilation from pion catalyzed multi-nucleon reactions during the expansion stage of heavy-ion collisions, because similar light nuclei numbers are obtained from these two approaches. 

Figure \ref{fig:yield} displays the collision centrality dependence of the yield ratio \hth~ for different values of $B_\Lambda$. The three collision centralities of 0-10\%, 10-30\%, and 30-50\% in this figure correspond to the charged-particle multiplicity $dN_{ch}/d\eta$ of about 1700, 1040, and 440, respectively, which are consistent with those measured in ALICE experiments~\cite{ALICE:2019hno}.  The yield ratio \hth~ is seen to be suppressed in more peripheral collisions (30-50\%) as compared to central collisions because of increasing emission source volume with  decreasing collision centrality. Also, the yield ratio \hth~ is more suppressed for a smaller $\Lambda$-separation energy or a larger $\Lambda-d$ distance with a variation by a factor of 2-3 for the different $\Lambda$-separation energies considered in the present study. The suppression of hypertriton production in non-central collisions is in accordance with earlier studies on light cluster production in collisions of small systems like $pp$ or $p$+Pb collisions at the LHC energies~\cite{Sun:2018mqq}. 

\begin{figure*}[!t]
    \centering
    \includegraphics[scale=1.45]{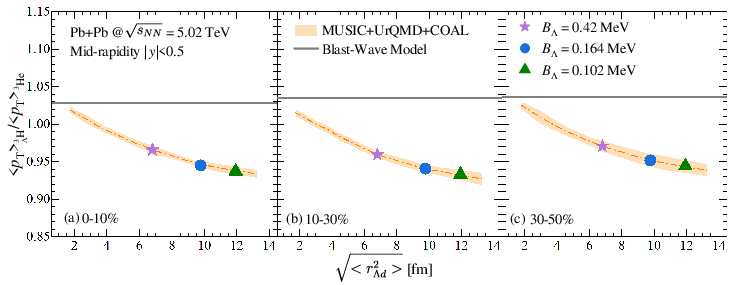}
    \caption{Ratio $\langle p_T\rangle_{^3_\Lambda\text{H}}/\langle p_T\rangle_{^3\text{He}}$ of hypertriton to helium-3 as a function of $\Lambda-d$ distance in Pb+Pb collisions at $\sqrt{s_{NN}} = 5.02$ TeV for 0-10\% (a), 10-30\% (b), and 30-50\% (c) centralities. The solid lines denote the prediction of the blast-wave model, whereas the shaded bands denote the prediction from the coalescence model. The solid symbols correspond to results obtained for various values of  $\Lambda$ separation energy $B_\Lambda$.}
    \label{fig:average_pt}
\end{figure*}

\emph{Softening of hypertriton $p_T$ spectrum}{\bf ---}
Figure \ref{fig:pt_spectra} displays by shaded bands the ratio \hth~as a function of transverse momentum ($p_T$) from the coalescence model. It is observed that  this spectrum ratio in all centralities also varies by a factor of 2-3 among the  three different values of the $\Lambda$ separation energy as for the yield ratio shown in Fig.~\ref{fig:yield}. Also, there is a pronounced decreasing trend in this ratio with an increase in $p_T$. This trend is in sharp contrast with the behavior predicted by the blast-wave (BW) model, shown by solid lines, which shows an increase of \hth~with increasing $p_T$. In the BW model, both  hypertriton and helium-3 are assumed in local thermal equilibrium at their kinetic freeze-out with an  invariant  momentum spectrum $f=E\text{d}^3N/\text{d}p^3$ given by~\cite{Cooper:1974mv, Schnedermann:1993ws}:
\begin{eqnarray}
f_\text{BL} \propto \int  m_T I_0\left(\frac{p_T\sinh(\rho)}{T_\text{kin}}\right) K_1\left(\frac{m_T\cosh(\rho)}{T_\text{kin}}\right)r \text{d}r. \notag\\
\label{eq:bw_pt}
\end{eqnarray}
In the above, $I_0$ and $K_1$ are the modified Bessel functions with their arguments depending on the transverse mass   $m_T=\sqrt{m^2+p_T^2}$,  the kinetic freeze-out temperature $T_{\rm kin}$, and the flow rapidity profile $\rho = \tanh^{-1}[(r/R_0)^n\beta_s]$ with   $r$ and $R_0$,    $\beta_s$, and $n$ being the radial distance in the transverse plane, the transverse radius,  the transverse expansion velocity at the surface, and the exponent of the velocity profile, respectively.  The blast-wave model can be viewed as a simple approximation to the hydrodynamic model used in describing heavy-ion collisions. The results shown in Fig.~\ref{fig:pt_spectra} for the blast-wave model are obtained by using the fitted kinetic freeze-out temperature ($T_\text{kin}$), flow velocity ($\beta_s$)  and exponent ($n$) in Ref.~\cite{ALICE:2019hno},  i.e.,  0.090 GeV, 0.907, and 0.735  for 0-10\% centrality, 0.094 GeV, 0.90, and 0.739 for 10-30\% centrality, and 0.105 GeV, 0.88, and 0.828 for 30-50\% centrality.   For  large $m_T$, $f_\text{BL}$ takes  the form  $f_\text{BL} \propto \text{exp}(-m_T/T_\text{eff})$ with $T_\text{eff}=T_{\rm kin}\sqrt{\frac{1+\langle\beta_T\rangle}{1-\langle\beta_T\rangle}}$ being the effective (blue-shifted) temperature. The ratio of the hypertriton  to the helium-3 spectrum at large $m_T$ is then given by 
\begin{eqnarray}
\frac{f_\text{BL}^\text{ht}}{f_\text{BL}^\text{he3}} \propto \text{exp}\left(-\frac{m^2_\text{ht}-m^2_\text{he3}}{(\sqrt{m^2_\text{ht}+p_T^2}+\sqrt{m^2_\text{he3}+p_T^2}) T_\text{eff}}\right),
\end{eqnarray}
which increases with increasing $p_T$ because the transverse flow drives the heavier hypertriton to higher momentum than the lighter helium-3, and this explains the increasing trend of the blast-wave model results (solid lines) shown in Fig.~\ref{fig:pt_spectra}. 

In the coalescence model, the ratio of hypertriton  to helium-3  $p_T$ spectra is, on the other hand, approximately given by~\cite{SchPRC59,Bellini:2018epz,Blum:2019suo}
\begin{eqnarray}
\frac{f_\text{COAL}^\text{ht}}{f_\text{COAL}^\text{he3}}\approx \frac{f_\text{BL}^\text{ht}}{f_\text{BL}^\text{he3}} 
\frac{[1+\frac{\sigma_\text{he3}^2}{2R^2(p_T)}]^{3}}{[1+\frac{\sigma_1^2}{2R^2(p_T)}]^{3/2}[1+\frac{\sigma_2^2}{2R^2(p_T)}]^{3/2}}, \label{eq:qs}
\end{eqnarray}
where $R(p_T)$ is the inhomogeneity Gaussian size of the nucleon and Lambda emission source for produced hypertrion and helium-3 of transverse momentum $p_T$, and the last factor (denoted by $\mathcal{C}$ hereafter) is from the quantum mechanical corrections~\cite{SchPRC59,Sun:2018mqq}  due to the hypertriton wave function. Because of the decreasing $R(p_T)$ with increasing $p_T$~\cite{ALICE:2015hvw} and  the relation $\sigma_2\gg \sigma_\text{he3}\sim \sigma_1$, the quantum correction factor $\mathcal{C}$ leads to a suppression of the \hyt~yield at larger $p_T$ compared to that of helium-3.  This can be explicitly seen as follows. Since $R(p_T)$ in the collisions considered in our study is larger than the sizes of hypertriton and helium-3, i.e., $R(p_T)\ge \sigma_2\gg\sigma_1,\sigma_{\rm he3}$, the correction factor can be approximately expressed as  
\begin{eqnarray}
\mathcal{C} \approx 1-\frac{3(\sigma_2^2+\sigma_1^2-2\sigma_{\text{he3}}^2)}{4R^2(p_T)}\approx 1-\frac{3\sigma_2^2}{4R^2(p_T)}, \label{fig:large}
\end{eqnarray}
which makes the ratio in Eq.~(\ref{eq:qs}) less than one and thus a large quantum correction.  With $R(p_T)$ decreasing with increasing $p_T$, the correction factor and thus the \hth~ratio decreases with increasing $p_T$ as shown in Fig.~\ref{fig:pt_spectra} for all collision centralities.  

For  a small emission source with $R(p_T)<\sigma_1,\sigma_{\rm he3}\ll\sigma_2$, the correction factor becomes approximately 
\begin{eqnarray}
&&\mathcal{C} \approx\frac{\sigma_{\text{he3}}^6}{\sigma_1^3\sigma_2^3}\left[1+3R^2(p_T)\left(\frac{2}{\sigma_{\text{he3}}^2}-\frac{1}{\sigma_{1}^2}-\frac{1}{\sigma_{2}^2}\right)\right]\notag\\
&&~\approx\frac{\sigma_{\text{he3}}^6}{\sigma_1^3\sigma_2^3}\left[1+\frac{3R^2(p_T)}{\sigma_{\text{he3}}^2}\left(2-\frac{\sigma_{\text{he3}}^2}{\sigma_{1}^2}\right)\right],
\end{eqnarray}
which also shows a suppression at large $p_T$ since the value of $R(p_T)$ decreases as  $p_T$ increases as a result of radial flow~\cite{ALICE:2011kmy,ALICE:2023zbh}. 

\emph{Suppression of hypertriton mean transverse momentum}{\bf ---} 
Shown in Fig.~\ref{fig:average_pt}   by shaded bands is the ratio $\langle p_T\rangle_{^3_\Lambda\text{H}}/\langle p_T\rangle_{^3\text{He}}$ of hypertriton to helium-3 mean-$p_T$ as a function of the $\Lambda-d$ distance, with the solid symbols corresponding to results obtained with various values of $B_\Lambda$. This mean transverse momentum ratio is seen to be less than one and  decreases  as the $\Lambda-d$ distance increases or the hypertriton binding energy $B_\Lambda$ decreases with a variation on the order of 10\%.   This variation is much smaller than the factor of 2-3 variation for the yield ratio in Fig.~\ref{fig:yield} and the spectrum ratio in Fig.~\ref{fig:pt_spectra} because of the weak dependence of the \hth~ratio on the transverse momentum. Also shown by solid lines is the prediction of the blast-wave model, which indicates that the value of the hypertriton to helium-3  mean-$p_T$ ratio has a slightly larger than unity value of about 1.03 for the 0-10\% collision centrality and about 1.04 for the 30-50\% collision centrality. The different hypertriton to helium-3 mean-$p_T$ ratio in the coalescence model and the blast-wave model is unlike the suppressed hypertriton production in peripheral collisions, which can be described by the coalescence model with the hypertriton wave function and also by the thermal model with a canonical correlation volume~\cite{Vovchenko:2018fiy}. Therefore, the observation of the softening of hypertriton transverse momentum spectrum and the reduction of its mean transverse momentum, albeit their small magnitude, is of great significance because it clearly contradicts to the thermal model predictions and can serve as a clean probe to the hypertriton production mechanism in relativistic heavy-ion collisions. 

Although a smaller mean $p_T$ for hypertriton than that for  helium-3 is seen for all collision centralities, the  centrality dependence of the softening effect shown in Fig.~\ref{fig:average_pt} is, however, rather weak because of the slowly decreasing radial flow with increasing collision centrality.   Since the radial flow is even smaller and the $p_T$ dependence of $R(p_T)$ is less prominent in collisions of small systems~\cite{ALICE:2011kmy,ALICE:2015hvw,ALICE:2023zbh},  the softening of hypertriton transverse momentum spectrum in $pp$ and $pA$ collisions is expected to  be even weaker in these collisions. For a quantitative study of the system size effect on the hypertriton transverse momentum spectrum requires, however, the extension of the present study to $pp$ and $pA$ collisions, which we leave for a future study.

\emph{Conclusion and outlook.}{\bf ---} In the present study, we have investigated hypertriton production in Pb+Pb collisions at $\sqrt{s_{NN}}=5.02$ TeV using the coalescence model with kinetic freeze-out nucleons and $\Lambda$ hyperons from a microscopic hybrid approach based on the MUSIC hydrodynamic model and the UrQMD hadronic transport model. We have found that the halo structure of hypertriton with a large $\Lambda-d$ distance of approximately $10$ fm leads to not only a suppression of \hyt ~yield but also a softening of its transverse momentum ($p_T$) spectrum with a weak centrality dependence. In particular, the mean $p_T$ of \hyt ~is found to be  smaller than that of helium-3  even in the most central collisions, which is in sharp contrast with the predictions of the blast-wave model. Such a quantum mechanical softening
of (anti-)hypertriton spectrum is a general feature and a natural outcome of the hypertriton wave function used in the final-state coalescence model, which can be readily tested in high-energy experiments with different beam energies and collision systems, providing thus the possibility to unravel the production mechanism of (anti-)hypernuclei in high-energy nuclear collisions and  to also obtain constraints on the $\Lambda$-nucleon interaction.
 
%\begin{acknowledgments} 
\emph{Acknowledgments.}{\bf ---} We thank Song Zhang and Bo Zhou for fruitful discussions and Chen Zhong for the maintenance of the CPU and GPU clusters. This work was supported in part by the  National Natural Science Foundation of China (Grants No. 12375121, No. 11891070, No. 11890714, No. 12147101, No. 12322508), the National Key Research and Development Project of China under Grant No. 2022YFA1602303, the STCSM (Grant No. 23590780100), the 111 Project,  and the U.S. Department of Energy under Award No. DE-SC0015266. 
%\end{acknowledgments} 

\begin{figure}[!t]
    \centering
    \includegraphics[scale=0.35]{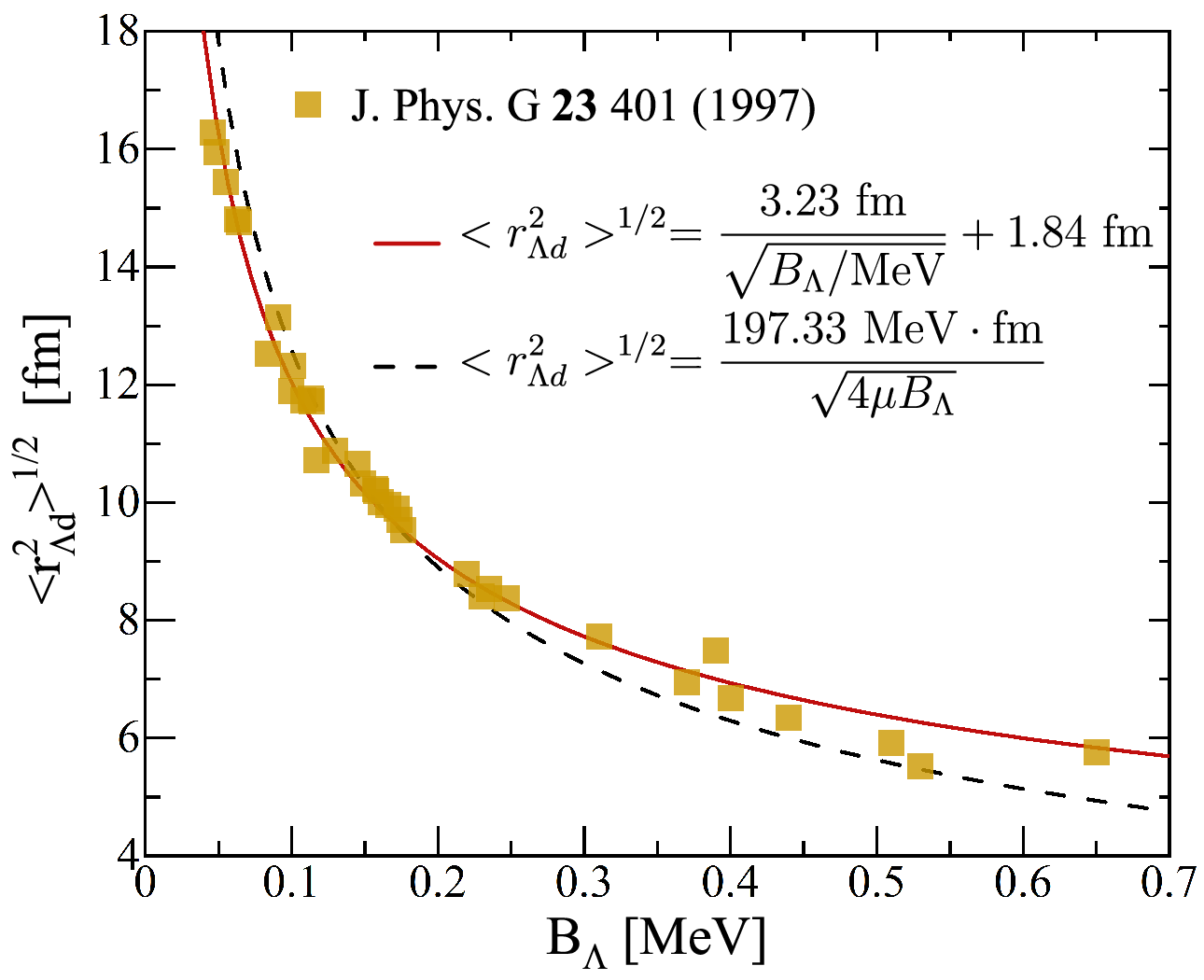}
    \caption{The $B_\Lambda$ dependence of $\Lambda$-d distance ($\sqrt{\langle r^2_{\Lambda d}\rangle}$) of hypertriton.  The solid squares denote theoretical results taken from Ref.~\cite {cobis_simplest_1997}. }
    \label{fig:binding_energy}
\end{figure}

\appendix
\section{APPENDIX}
\label{appendix}
\emph{$\Lambda$ separation energy and $\Lambda-d$ distance.}{\bf ---} Figure~\ref{fig:binding_energy} depicts the $\Lambda-d$ distance $\sqrt{\langle r^2_{\Lambda d}\rangle}$  of hypertriton as a function of $\Lambda$ separation energy $B_\Lambda$. The symbols in Fig.~\ref{fig:binding_energy} are theoretical results taken from Ref.~\cite {cobis_simplest_1997}.  Because the hypertriton is a loosely bound state of a deuteron and a $\Lambda$ hyperon, one has the relation (black dashed line of Fig.~\ref{fig:binding_energy})
\begin{equation}
\sqrt{\langle r_{\Lambda d}^{2} \rangle } \approx \frac{197.33~\text{MeV}\cdot \text{fm}}{\sqrt{4\mu B_{\Lambda }}} ,\label{lamD}
\end{equation}
where $\mu =m_{\Lambda } m_{d} /(m_{\Lambda } +m_{d} )=700$ MeV is the reduced mass. Instead of  Eq.~(\ref{lamD}), we use the following parametrization for  the $\Lambda-d$ distance  (red solid line in Fig.~\ref{fig:binding_energy}),
\begin{equation}
{\displaystyle \sqrt{\langle r_{\Lambda d}^{2} \rangle } \approx \frac{a}{\sqrt{B_{\Lambda }/\text{MeV}}} +b,}
\end{equation}
with parameters $a=3.226\pm 0.062$ fm and  $b=1.836\pm 0.173$ fm. For $B_\Lambda$=0.42  MeV (STAR), 0.164 MeV (world average), and 0.102 MeV (ALICE), the corresponding values of $\sqrt{\langle r_{\Lambda d}^{2}\rangle}$ are 6.8 fm, 9.5 fm, and 11.9 fm, respectively. The size parameter $\sigma_2$ in the coalescence model is related to the $\Lambda-d$ distance by $\sigma_2=\frac{2}{3}\sqrt{\langle r_{\Lambda d}^2\rangle}$, and the values of $\sigma_2$ are 5.45 fm, 6.52 fm, and 7.96 fm, respectively, for the above three values of $B_\Lambda$.

%\bibliography{ref}


 \end{document}